\documentclass[jgrga]{AGUTeX}
\usepackage{color}
\usepackage{graphicx}
\authorrunninghead{MASAHIRO HOSHINO}

\titlerunninghead{MHD Instability in Reconnection Jets}

\authoraddr{Corresponding author: M. Hoshino,
Faculty of Science, The University of Tokyo, Tokyo 113-0033, Japan
(hoshino@eps.s.u-tokyo.ac.jp)}

\begin{document}
%
%
\title{Generation of Alfv\'enic Waves and Turbulence in Reconnection Jets}
%
%
\authors{Masahiro Hoshino,\altaffilmark{1}
Katsuaki Higashimori,\altaffilmark{1}}
\altaffiltext{1}{Department of Earth and Planetary Science,
The University of Tokyo, Tokyo, Japan}
%
%
\begin{abstract}
The magnetohydrodynamic linear stability with the localized 
bulk flow oriented parallel to the neutral sheet is investigated, 
by including the Hall effect and the guide magnetic field.
We observe three different unstable modes: a ``streaming tearing'' mode 
at a slow flow speed, a ``streaming sausage'' mode at a medium flow speed, 
and a ``streaming kink'' mode at a fast flow speed. The streaming tearing 
and sausage modes have a standard tearing mode-like structure with 
symmetric density fluctuations in the neutral sheet, while the kink 
mode has an asymmetric fluctuation. The growth rate of the streaming tearing 
mode decreases with increasing magnetic Reynolds number, while the growth rates 
of the sausage and kink modes do not depend strongly on the Reynolds number. 
The sausage and kink modes can be unstable for not only 
super-Alfv\'enic flow but also sub-Alfv\'enic flow when the lobe density is low.  
The wavelengths of these unstable modes are of the same order of 
magnitude as the thickness of the plasma sheet.  Their maximum growth 
rates are higher than that of a standard tearing mode,
and under a strong guide magnetic field, the growth rates of the 
sausage and kink modes are enhanced, while under a weak guide magnetic 
field, they are suppressed.  For a thin plasma sheet with the Hall effect, 
the fluctuations of the streaming modes can exist over the plasma sheet.  
These unstable modes may be regarded as being one of the processes 
generating Alfv\'enic turbulence in the plasma sheet during magnetic 
reconnection.
\end{abstract}
%
%
\begin{article}
\section{Introduction}
Alfv\'enic fluctuations and turbulence with orders of magnitude from 
several tens $R_E$ down to ion inertia lengths are often observed in 
the earth's magnetotail \citep[e.g.,][]{Russell72,Hoshino94,Bauer95,Zelenyi14}.  
The amplitude of these wave fluctuations in the plasma sheet can reach 
up to half the value of the lobe magnetic field, and these waves can carry 
a significant fraction of the plasma energy in the plasma sheet 
\citep[e.g.,][]{Borovsky97}. It has been discussed that turbulence with 
a large wave energy density plays an important role in the transport of 
mass and momentum in the magnetotail 
\citep[e.g.,][]{Borovsky03,Zimbardo10}. Understanding the effects of this 
turbulence is believed to be a key factor in obtaining an overall 
understanding of the various plasma phenomena that occur both in space, 
and also in astrophysical plasmas \citep[e.g.,][]{Birn12,Lazarian12}.

To date, many researchers have discussed the importance of turbulence in 
the plasma sheet in the context of magnetic reconnection.
It has been argued that turbulence plays a crucial role in the dynamic 
evolution of magnetic reconnection and that the magnetic energy dissipation 
rate can be enhanced by turbulence 
\citep[e.g.,][]{Matthaeus86,Lazarian99,Loureiro09,Higashimori13,Yokoi13}.
In addition to its role in the dynamic evolution of reconnection, turbulence 
is believed to play an essential role not only in thermal plasma heating 
but also in nonthermal particle production through the stochastic scattering 
of particles \citep[e.g.,][]{Veltri98,Greco02,Zelenyi98,Zelenyi11,Lazarian12}.
However, in previous studies, it was postulated that turbulence can be generated 
in the high $\beta$ plasma sheet with a high magnetic Reynolds number, but 
the detailed mechanism required to generate Alfv\'enic waves and turbulence 
during the dynamic evolution of reconnection is not understood as yet.

Several possible candidates for generating Alfv\'enic fluctuations in the 
plasma sheet have been proposed: (1) the anisotropic ion beams observed in 
the plasma sheet boundary layer (PSBL) can generate magnetohydrodynamic (MHD) 
waves through a family of ion--ion beam instabilities 
\citep[e.g.,][]{Gary91,KraussVerban95,Grigorenko11}. 
The excited Alfv\'enic wave in the boundary may penetrate into the plasma 
sheet because the refractive index of the Alfv\'en wave in the plasma 
sheet is higher than in the lobe plasma region.
(2) An elongated magnetic diffusion region with strong electric 
currents may make the tearing mode unstable, and many small-scale plasmoids 
created by the tearing mode instability can emanate from the 
diffusion region 
\citep[e.g.,][]{Loureiro07,Samtaney09,Bhattacharjee09,Pucci14}. 
(3) The magnetic field pile-up region, where the reconnection jet stops suddenly,
is expected to excite various plasma waves by releasing the bulk flow energy
\citep[e.g.,][]{Hoshino98,Hoshino01}. 
In dawn-dusk direction parallel to the electric current, the interchange 
instability can be excited, because the gradient of the plasma density is 
opposite to the decelerating plasma flow profile.
\citep[e.g.,][]{Nakamura02,Lapenta11}. 
(4) Turbulence in the solar wind may penetrate into the magnetotail across 
the magnetopause. Even if the magnetosphere is encircled by a closed 
magnetic field line, fluctuations in the magnetosonic waves can propagate 
perpendicular to the magnetic field. (5) In addition to the above possible
origins of Alfv\'enic fluctuations, a reconnection jet at Alfv\'enic 
speed can be a free energy source for the MHD instability from the release of 
bulk flow energy.

In this paper, we focus on an MHD instability triggered by a bulk flow 
plasma in the reconnection downstream, and propagating to the parallel
to the bulk flow.  In this situation, 
some reconnection simulations using the hybrid codes suggested 
the generation of a kink-type motion of the reconnection exhaust as one 
of possible origins of turbulence in the plasma sheet 
\citep{Lottermoser98,Arzner01,Higashimori12,Liu12}, and a tearing mode
MHD simulation under the localized bulk flow claimed the rapid excitation 
of the sausage-type fluctuations \citep{Sato82}.
In addition to these nonlinear simulation results, the MHD linear instability 
with the localized bulk flow oriented parallel to the neutral sheet, which focuses
on only the reconnection exhaust, has been extensively 
investigated by \citet{Shigeta85,Wang88,Lee88,Biskamp98}.

In the MHD linear stability analysis, three important modes have been discussed.
One of the unstable modes is the ``streaming tearing'' mode \citep{Shigeta85,Wang88}, 
categorized as a resistive MHD instability, and a second mode 
is the ``streaming sausage'' mode \citep{Lee88,Biskamp98}, 
categorized as an ideal MHD instability. 
The eigenfunctions of the streaming tearing/sausage modes are similar to 
the standard tearing mode \citep{FKR63}, whose density perturbation is 
symmetrical to the neutral sheet. The streaming tearing/sausage modes are 
not zero-frequency modes, and the tearing islands drift with the bulk flow speed.
In addition to the two symmetric unstable modes above, 
an asymmetric unstable mode can be observed, 
which we call the ``streaming kink'' mode \citep{Lee88,Biskamp98}.
This mode can be categorized as an ideal MHD instability, and 
has an instability mechanism that is similar to the streaming sausage mode.
(Note that the terminology of the drift-kink instability
whose mode is propagating along the electric current direction is different 
from the streaming kink instability discussed in this paper
\citep{Pritchett96,Zenitani05,Fujimoto11}.)

In this paper, using the standard method of matrix eigenvalue analysis, 
we studied the above three different types of streaming modes
from sub-Alfv\'enic to super-Alfv\'enic regime, 
by paying special attention to the earth's 
magnetotail where the lobe density is low, 
because the dependence of bulk flow speed on the linear growth rate 
has not been systematically investigated, and because the dilute lobe 
plasma environment has not been taken into account. 
We observed that the streaming sausage and streaming kink modes 
became unstable in the magnetotail when the bulk flow speed 
reaches a value of several tens of percent of the Alfv\'en speed.
More importantly, we studied the streaming modes under the effects of the Hall current 
and the guide magnetic field, and propose a possible origin of turbulence in the plasma 
sheet during magnetic reconnection. 
In Section 2, the basic model is described with
a bulk flow plasma in the center of the plasma sheet, and 
we introduce the method used to solve the linear stability of the streaming plasmas.
In Section 3, we discuss our linear stability results for both
the streaming tearing/sausage modes and the streaming kink mode. In Section 4, we
summarize our results and provide a perspective on the streaming 
instabilities.

\section{Streaming Plasma Sheet Model and Linear Analysis}
In this paper, we discuss the streaming MHD modes in the framework of
the Hall MHD. As discussed later, the streaming instability itself 
can be activated in the standard MHD regime without the Hall effect,
but we use the Hall MHD equation having its application to the earth's 
magnetotail in mind, because from 
satellite observations, the thickness of the reconnection 
region is known to be of the order of the ion inertial length 
\citep[e.g.,][]{Sergeev93,Asano03,Nakamura06}.

We studied the linear stability of a Harris-type plasma sheet with 
a finite bulk flow oriented parallel to an antiparallel magnetic field.
The initial equilibrium state was set to the Harris solution, where
the magnetic field $B_x$ is given by
\begin{equation}
B_x= B_{\rm lobe} \tanh(y/\lambda),
\end{equation}
and $B_y = 0$ and $B_z = {\rm const.}$, where $\lambda$ is the thickness of the plasma sheet. 
The plasma density, $\rho$, is expressed by
\begin{equation}
\rho=\frac{\rho_{\rm ps}}{\cosh^2(y/\lambda)}+\rho_{\rm lobe}.
\end{equation}
In this paper, we assumed a finite background density, $\rho_{\rm lobe}$, exists over the 
entire plasma sheet. The temperature for the uniform component, 
$\rho_{\rm lobe}$, was assumed to be $T_{\rm lobe}=0$.

In addition to the standard Harris state, we added a finite bulk flow,
whose velocity profile, $v_x$, is given by
\begin{equation}
v_x=\frac{v_{\rm jet}}{\cosh^2(y/\lambda_{\rm jet})},
\end{equation}
and $v_y=v_z=0$, where $\lambda_{\rm jet}$ is the thickness of the bulk flow. 
The bulk flow may mimic the reconnection exhaust with an Alfv\'enic jet.
An MHD instability coupled with the 
streaming plasma may be activated because of this streaming plasma.

We used a set of compressible MHD equations
\begin{equation}
\frac{\partial \rho}{\partial t} = -\nabla \cdot (\rho \vec{v}), 
\end{equation}
\begin{equation}
\frac{\partial \vec{v}}{\partial t}+ \vec{v} \cdot \nabla \vec{v} =
-\frac{1}{\rho} \nabla p + \frac{1}{\rho c} \vec{j} \times \vec{B}
+ \nu \nabla^2 \vec{v}, 
\end{equation}
\begin{equation}
\frac{\partial \vec{B}}{\partial t} = \nabla \times (\vec{v} \times \vec{B}
- \frac{1}{en} \vec{j} \times \vec{B})+ \eta \nabla^2 \vec{B}, 
\end{equation}
where the terms $e$ and $n$ in the $\vec{j} \times \vec{B}$ Hall component represent 
the electric charge and the number density, respectively. 
The equation of state is assumed to be adiabatic, i.e., 
$p \propto \rho^{\gamma}$, where the ratio of the specific heat is 
$\gamma=5/3$. The electron temperature was set to zero.
The terms $\nu$ and $\eta$ are the fluid viscosity and the electric resistivity, respectively.

After linearizing the above equations using the standard perturbation method 
in $x$ and $y$ two-dimensional space, we solved the set of linearized 
equations using the matrix eigenvalue method, which is often used in 
the analysis of plasma instabilities \citep[e.g.,][]{Hoshino91}.
The time derivative, $\partial f(x,y,t)/\partial t$, of the physical quantity, $f$, 
and the spatial derivative along 
the plasma sheet, $\partial f(x,y,t)/\partial x$, can be expressed by 
$-i \omega \tilde{f}(k,y,\omega)$ and $i k \tilde{f}(k,y,\omega)$ using 
a Fourier--Laplace transformation in time and space, respectively.
In this paper, we assumed that the physical quantities were uniform in the 
$z$ direction, i.e., $\partial/\partial z =0$.

The spatial derivatives in the direction vertical to the plasma sheet, 
$\partial f/\partial y$, can be 
approximated by the fourth-order, finite difference representation of
the differentiation with respect to $y$.
In addition to this, we used a nonuniform grid spacing. For example, 
the first derivative in the finite difference representation 
$\partial f/\partial y$ can be expressed by
\begin{equation}
\frac{\partial f(x,y_i,t)}{\partial y}=\sum_{j=-2}^{2}a_{j}f(x,y_{i+j},t),
\end{equation} 
where $y_i$ is the position of $i$-th grid, and
\begin{displaymath}
a_{i\pm2} = \pm \frac{\Delta_{i\pm1}\Delta_{i\mp1}\Delta_{i\mp2}}
{\Delta_{i\pm2}
(\Delta_{i\mp2}+\Delta_{i\pm2})
(\Delta_{i\pm1}-\Delta_{i\pm2})
(\Delta_{i\mp1}+\Delta_{i\pm2})}, 
\end{displaymath}
\begin{displaymath}
a_{i\pm1} = \pm\frac{\Delta_{i\pm2}\Delta_{i\mp1}\Delta_{i\mp2}}
{\Delta_{i\pm1}
(\Delta_{i\mp1}+\Delta_{i\pm1})
(\Delta_{i\mp2}+\Delta_{i\pm1})
(\Delta_{i\pm2}-\Delta_{i\pm1})}, 
\end{displaymath}
\begin{displaymath}
a_{i} = \frac{\Delta_{i-1}+\Delta_{i-2}}{\Delta_{i-1}\Delta_{i-2}}-
\frac{\Delta_{i+1}+\Delta_{i+2}}{\Delta_{i+1}\Delta_{i+2}}, 
\end{displaymath}
where $\Delta_{i\pm1}=|y_{i\pm1}-y_{i}|$ and 
$\Delta_{i\pm2}=|y_{i\pm2}-y_{i}|$. 
A nonuniform grid size is useful for resolving the resistive layer 
in the vicinity of the neutral sheet for the case of a large magnetic Reynolds number
\citep[e.g.,][]{Hoshino91}.
The grid size in the vicinity of the neutral sheet was set to be 
five to 50 times smaller than that in the lobe region. 
The number of grid points was chosen to be in the range 
$400$ to $1600$ to ensure the convergence of the numerical results.

The size of the plasma sheet, $|y|<L$, was set to be $L/\lambda=5$,
and the boundary condition was assumed to be $f(x,|y|=L,t)=0$ 
for the spontaneous reconnection model.
In our study, we used the magnetic Reynolds number, 
$R_M=V_A \lambda/\eta = 10^3$, and the fluid Reynolds number, 
$R_V=V_A \lambda/\nu = 10^4$, in most of our calculations, except
for the cases mentioned. The other plasma parameters discussed in this paper are
 listed in Table 1.

\section{Results of the Streaming MHD Instability}
\subsection{Streaming instability with symmetric or \\
asymmetric perturbation}
\begin{figure}
\centering
\includegraphics[width=\textwidth]{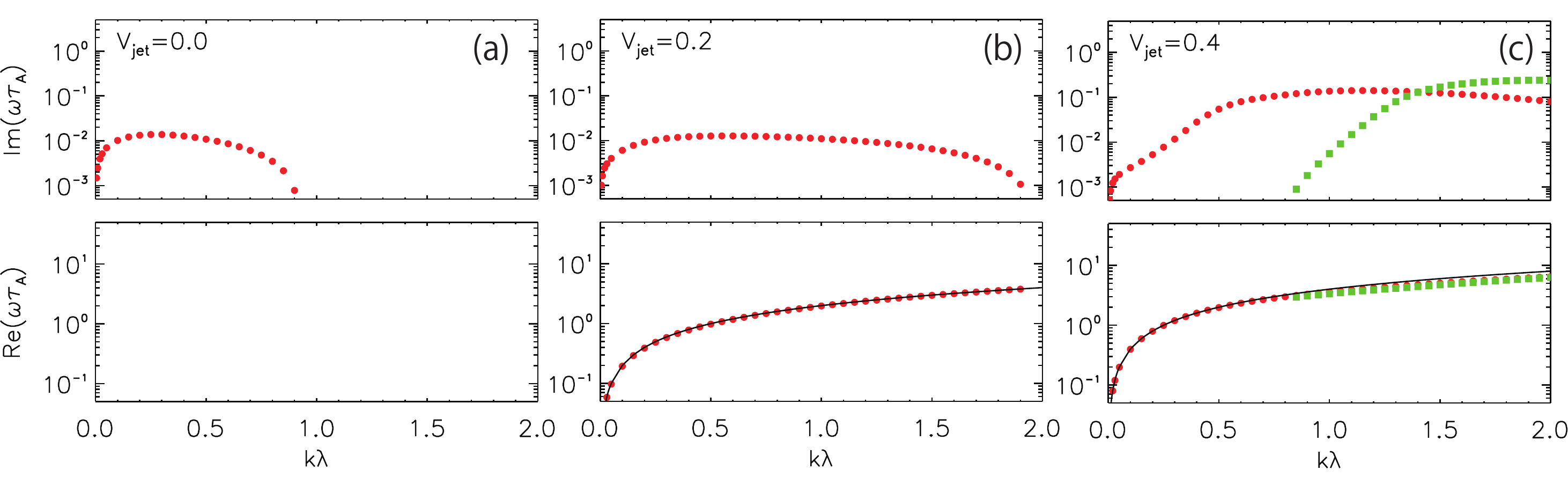}
\caption{Linear growth rate, Im$(\omega \tau_A)$, (top) and oscillation 
Frequency, Re$(\omega \tau_A)$, (bottom) as the function of wave number, 
$k \lambda$, for three different bulk flow speeds: 
(a) $V_{jet}=v_{\rm jet}/V_{\rm A,lobe}=0$, i.e., no bulk flow, 
(b) $V_{jet}=0.2$, and (c) $V_{jet}=0.4$.
The solid black lines in the bottom panels (b) and (c) are the Doppler
shift frequencies of Re$(\omega \tau_A)=(k \lambda) V_{\rm jet}$.}
\label{fig_Vjet}
\end{figure}

Figures \ref{fig_Vjet}a--c show the linear growth rates 
obtained from our matrix eigenvalue analysis for three different bulk 
flow speeds. The top of each panel 
shows the linear growth rate, $\rm{Im}(\omega \tau_A)$, 
while the bottom of each panel shows the oscillation frequency, $\rm{Re}(\omega \tau_A)$,
where $\tau_A=\lambda/V_A$ is the Alfv\'en transit time. 
Figure \ref{fig_Vjet}a shows the case where there was no bulk flow, 
$V_{\rm jet}=0$, and we only obtained the standard tearing mode \citep{FKR63}, 
where the unstable region only appeared for 
$k \lambda < 1$, and the oscillation frequency was zero. 

Figure \ref{fig_Vjet}b shows the case for a finite bulk flow speed 
with $V_{\rm jet}=v_{\rm jet}/V_{\rm A,lobe}=0.2$, normalized to the lobe 
Alfv\'en speed, defined by $V_{\rm A,lobe}=B_{\rm lobe}/\sqrt{4 \pi \rho_{\rm lobe}}$.
Note that the bulk flow speed is normalized to the lobe Alfv\'en speed, while
the Alfv\'en transit time was normalized to the Alfv\'en speed, defined by
the total plasma density in the neutral sheet, 
$\rho_0=\rho_{\rm ps}+\rho_{\rm lobe}$, because the reconnection jet speed 
was expected to reach to the lobe Alfv\'en speed when a switch-off shock 
formed in the boundary.

In Figure \ref{fig_Vjet}b, it can be seen that the unstable region extended to 
the larger wave number region for $k \lambda > 1$, and that the growth rate was 
enhanced slightly compared with the standard tearing mode in Figure \ref{fig_Vjet}a. 
The oscillation frequency, $\rm{Re}(\omega \tau_A)$, denoted by
the red-colored circles can be approximated by the drift frequency/Doppler shift 
frequency of $\rm{Re}(\omega \tau_A) \simeq (k \lambda) V_{\rm jet}$, 
depicted by the black curve.
We confirmed that the behavior of the linear growth rate was basically the 
same as the ``streaming tearing'' mode studied by \citet{Shigeta85,Wang88}, 
and we found that from the eigenfunctions the density perturbation was symmetric to the neutral 
sheet (not shown here).

Figure \ref{fig_Vjet}c shows the growth rate and the oscillation frequency 
for a higher bulk flow speed with $V_{\rm jet}=v_{\rm jet}/V_{\rm A,lobe}=0.4$. 
In addition to the streaming tearing mode denoted by the red-colored 
circles, we observed the streaming kink mode, denoted by the green-colored squares, 
whose density perturbation, $\rho$, and vector potential, $A_z$, became 
asymmetric versus the neutral sheet. The streaming kink mode could not be
excited below a given minimum threshold speed. However, since 
the threshold speed was less than the lobe Alfv\'en velocity, 
$V_{\rm A,lobe}$, the streaming kink mode can be easily generated during magnetic
reconnection.
Another interesting point is that the growth rate of the streaming kink 
mode was larger than the streaming tearing mode, and the streaming kink mode
may dominate the plasma sheet turbulence in the reconnection jet.

\begin{figure}
\centering
\includegraphics[width=40pc]{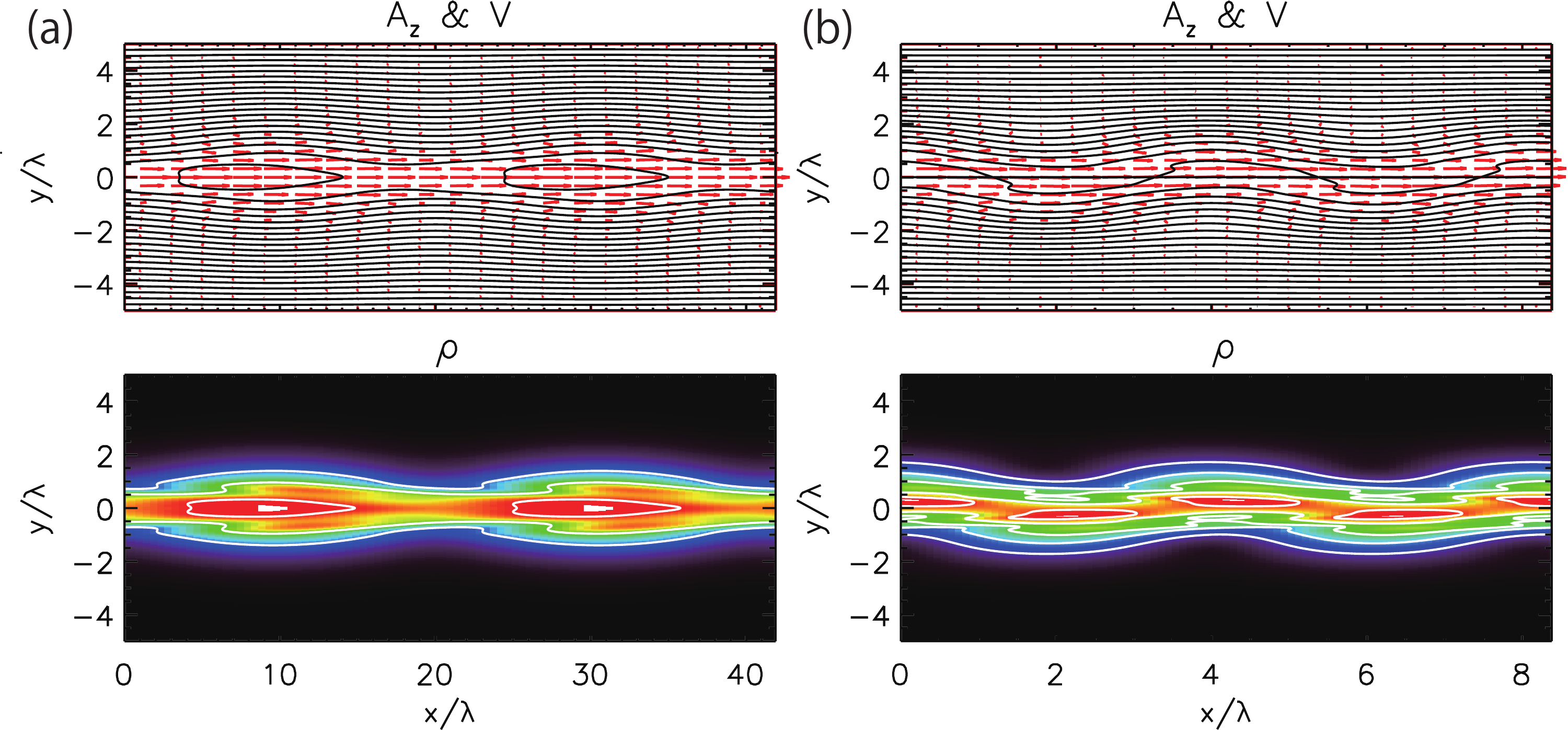}
\caption{Unstable structures for the vector potential, $A_z$, (top)
and the plasma density, $\rho$, (bottom), which were reconstructed from 
the eigenfunctions and eigenvalues. The flow vectors (red-colored arrows) are 
superposed on the magnetic field lines (black) of the 
contours for $A_z$. (a) The streaming tearing mode with 
$V_{jet}=0.2$ and $k \lambda=0.3$ in Figure \ref{fig_Vjet}b, 
and (b) the streaming kink mode with 
$V_{jet}=0.4$ and $k \lambda=1.5$ in Figure \ref{fig_Vjet}c.}
\label{fig_Egn}
\end{figure}

Figure \ref{fig_Egn} shows typical linear structures of the 
vector potential, $A_z$, and plasma density, $\rho$, in $(x,y)$ space 
that were reconstructed from the eigenvalues and eigenfunctions 
in Fourier $(k,y)$ space. The first-order perturbations 
were superposed onto the zero-order quantities, and we assumed
the magnitude of the first-order peak density was $30 \%$ 
of the zero-order density in the neutral sheet. 
Figure \ref{fig_Egn}a shows the streaming tearing mode for 
$(V_{\rm jet},k \lambda)=(0.2,0.3)$; Figure \ref{fig_Egn} b shows 
the streaming kink mode for
$(V_{\rm jet},k \lambda)=(0.4,1.5)$. The structure of the 
streaming tearing mode was similar to that of the standard tearing mode,
except for the magnetic islands that streamed with the zero-order
bulk flow speed, $v_{\rm jet}$. On the other hand, the structure of 
the streaming kink mode had a periodic distortion in the neutral 
sheet.

\subsection{Maximum growth rate}
So far, we have discussed that the streaming tearing mode is unstable for 
a relatively slow bulk flow regime, while the streaming kink mode 
appears to be unstable in a relatively fast flow regime. Let us now study the
behavior of the two unstable modes as a function of the bulk flow 
speed, $V_{\rm jet}$.

Figure \ref{fig_Max} shows the maximum growth rate and 
the corresponding wave number as a function of the bulk flow speed, 
$V_{\rm jet}$. In our numerical calculations, we surveyed the growth rate
and the oscillation frequency for a given wave number for 
$k\lambda=0.05 \times i$, for $i=0$ to $63$. The wave number and 
bulk flow were normalized to the thickness of the plasma sheet, 
$\lambda$, and the lobe Alfv\'en velocity, $V_{\rm A,lobe}$, respectively. 
The red-colored circles denote the streaming tearing mode, 
while the green-colored circles denote the streaming kink mode.
We observed that the streaming tearing mode with a symmetric 
density perturbation dominated the regime where $V_{\rm jet} < 0.4$, 
while the growth rate of the streaming kink mode became larger than that of
the streaming tearing mode when the bulk flow speed exceeded about 
$V_{\rm jet} > 0.4$. These modes have a 
long wavelength nature for $k \lambda < 2$.

By carefully examining the streaming tearing mode denoted by the 
red-colored circles, we could observe that the characteristics of the unstable wavelengths were 
different below and above a value of $V_{\rm jet} \sim 0.25$. The unstable 
wavelengths for $V_{\rm jet}<0.25$ were the long wavelengths with 
$k \lambda < 1$, while for $V_{\rm jet}>0.25$, the unstable wavelengths 
occurred for $1 < k \lambda < 2$. Moreover, the growth curve of 
${\rm Im}(\omega \tau_A)$ showed a discontinuous transition from 
Im$(\omega \tau_A) \sim 10^{-2}$ to $5 \times 10^{-2}$ around a value of
$V_{\rm jet} \sim 0.25$. This suggests that two different modes 
may exist below and above $V_{\rm jet} \sim 0.25$. 

As we discuss this behavior further in the next subsection, 
\ref{sec:ReynoldsNumber}, it can be approximated that the linear growth rate of 
the symmetric unstable mode for $V_{\rm jet} < 0.25$
strongly depends on the magnetic Reynolds number, while the linear growth rate of 
the symmetric unstable mode for 
$V_{\rm jet} > 0.25$ does not depend on the magnetic Reynolds number. 
Therefore, we distinguish the symmetric perturbations from their 
dependence of the linear 
growth rate on the magnetic Reynolds number, $R_M$. 
One of these modes is the ``streaming tearing'' mode, categorized as a 
resistive MHD mode \citep{Shigeta85,Wang88},
and the other mode is the ``streaming sausage'' mode, classified as 
an ideal MHD mode \citep{Lee88,Biskamp98}. 
The streaming tearing and streaming sausage modes are observed 
in relatively slow and fast bulk flow speeds, respectively.

The perturbed structure of the streaming sausage mode is similar
to that of the streaming tearing mode (not shown here), and the vector potential, 
$A_z$, and the plasma density, $\rho$, for the streaming tearing/sausage 
modes show similar properties to the standard tearing mode, namely 
those perturbations are symmetric against the neutral sheet.

\subsection{Reynolds number dependence, $R_M$}
\label{sec:ReynoldsNumber}
\begin{figure}
\centering
\includegraphics[width=20pc]{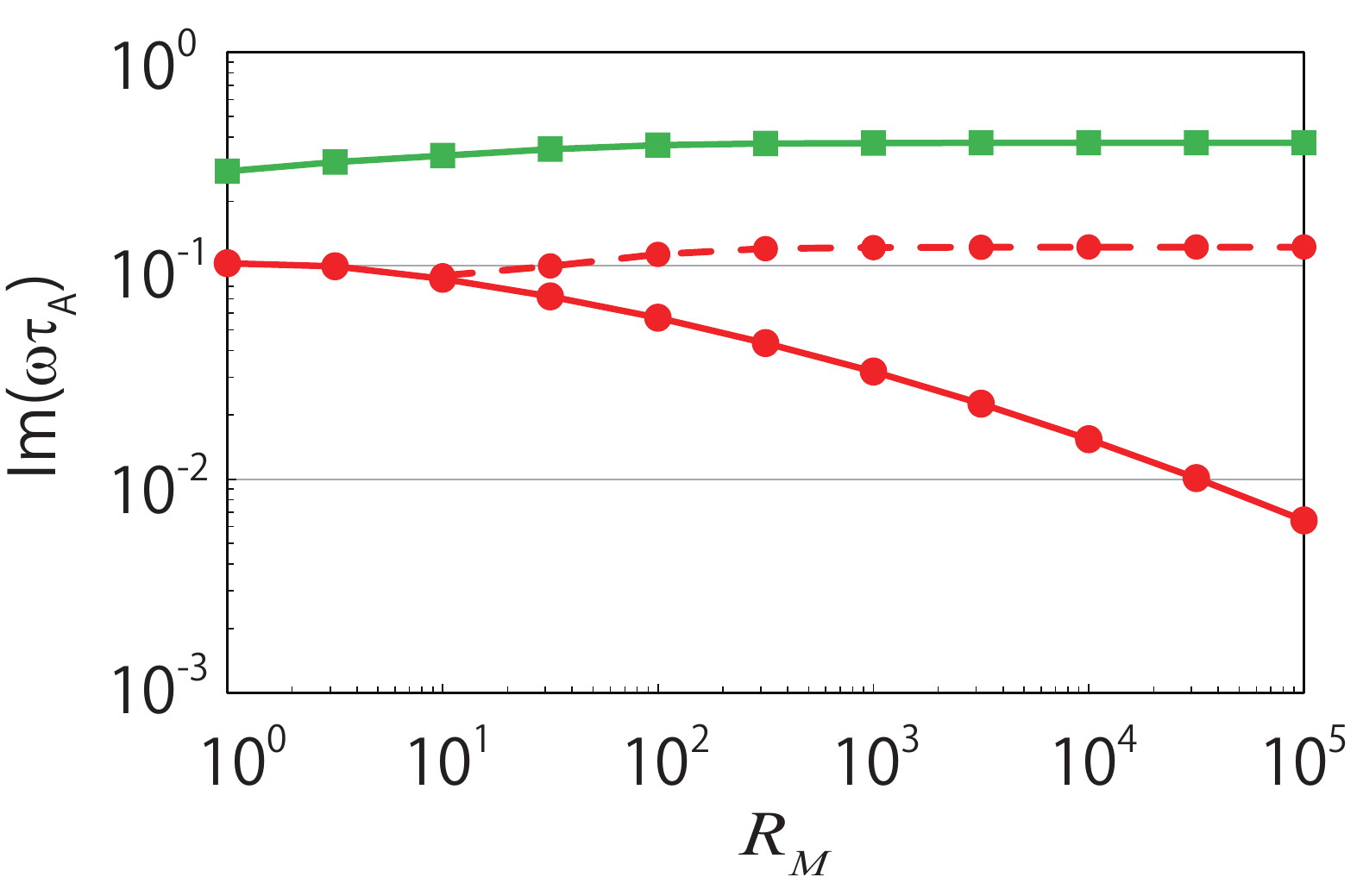}
\caption{Growth rates, Im$(\omega \tau_A)$, for the streaming tearing mode 
for $k \lambda=0.3$ (red-colored solid line with circles), the streaming sausage 
mode for $k \lambda=1.0$ (red-colored dashed line with circles), and the streaming 
kink mode for $k \lambda=1.0$ (green-colored solid line with squares) as a 
function of magnetic Reynolds number, $R_M$. The bulk flow speed was set 
to be $V_{\rm jet}=0.5$ for all modes.}
\label{fig_Rm}
\end{figure}

Figure \ref{fig_Rm} shows the linear growth rates for the streaming tearing,
streaming sausage, and streaming kink modes as a function of the magnetic Reynolds number. 
The bulk flow speed was fixed to be $V_{\rm jet}=0.5$ for all cases. 
The red-colored circles with the solid line and the red-colored circles with the dashed line
denote the growth rates of the symmetric density perturbation for
the streaming tearing mode with $k \lambda=0.3$ and the streaming
sausage mode with $k \lambda=1.0$, respectively.
The green-colored squares correspond to the streaming kink mode with an
asymmetric density perturbation for the wave number with $k \lambda=1$. 
There may be more than one unstable mode for any given wave number. 
The streaming sausage and streaming kink modes showed 
symmetric and asymmetric density fluctuations, respectively.

The growth rate of the streaming tearing mode decreased with increasing 
magnetic Reynolds number, $R_{M}$, and the growth rate was approximately 
Im$(\omega \tau_A) \propto R_{M}^{-1/3}$, which showed a similar 
dependence to that of the standard tearing mode of $R_M^{-3/5}$ \citep{FKR63}, 
in the sense that it was a decreasing function.  The streaming tearing mode
is slightly enhanced for a large magnetic Reynolds number, by utilizing
the free energy of the streaming bulk flow.
On the other hand, the growth rate of the streaming sausage and streaming kink 
modes did not depend strongly on the Reynolds number, and the growth 
rates were almost constant in the high magnetic Reynolds 
number regime. 
This result is consistent with the streaming sausage and 
streaming kink modes under an ideal MHD regime discussed by 
\citet{Lee88,Biskamp98}. 
Therefore, in a plasma 
medium with a high magnetic Reynolds number, $R_M$, the streaming sausage 
and streaming kink modes may play an important role in the excitation of 
Alfv\'enic fluctuations, even if the streaming tearing mode cannot
be excited.

\subsection{Lobe density dependence, $\rho_{\rm lobe}/\rho_0$}
We confirmed that the streaming tearing/sausage 
and streaming kink instabilities 
can be excited by adding a bulk flow plasma to the plasma sheet
\citep[e.g.,][]{Shigeta85,Wang88,Lee88,Biskamp98}, 
and these instabilities are thought to be important in the 
reconnection jet region. Since the reconnection jet speed is known to 
be the lobe Alfv\'en velocity if a switch-off slow mode shock
is formed, the generation of turbulence by the streaming instabilities 
may be strongly controlled by the lobe plasma density.

To study the dependence of the lobe plasma density on the linear
stability, the data in Figure \ref{fig_LobeN} show the growth rate 
for $\rho_{\rm lobe}/\rho_0=0.01$, $0.05$, and $0.1$, where 
$\rho_0=\rho_{\rm ps}+\rho_{\rm lobe}$ is the total plasma density 
in the neutral sheet \citep[e.g.,][]{Ishisaka01}. Figure \ref{fig_LobeN}a 
shows the same data as in Figure \ref{fig_Max}.
It can be seen that the streaming tearing and streaming sausage modes with a
symmetric perturbation are excited in the relatively slow bulk flow
regime, while the streaming kink mode with an asymmetric 
perturbation appears in a relatively fast flow regime. The onset 
speeds of the streaming sausage and kink modes 
increased with increasing lobe plasma 
density, $\rho_{\rm lobe}/\rho_0$.

Discontinuous features in the growth rates of the symmetric 
perturbation, suggesting a transition from the streaming
tearing mode to the streaming sausage mode, were observed around 
$V_{\rm jet} \sim 0.25$, $0.55$, and $0.75$ for 
$\rho_{\rm lobe}/\rho_0=0.01$, $0.05$, and $0.1$, respectively.
The onset speed of the transition also increased with increasing 
the lobe plasma density. The growth rates for the streaming
tearing and streaming sausage modes did not change appreciably on increasing 
the lobe plasma density, but growth rates for the streaming kink mode could be 
suppressed because of the contribution of the dense lobe plasma.

Let us now look at the detail on the streaming tearing mode.
We can recognize modest peaks in the growth rates for relatively small 
jet velocities, namely the positions of the peaks are
$V_{\rm jet} \sim 0.1$ for $\rho_{\rm lobe}/\rho_0=0.01$,
$V_{\rm jet} \sim 0.2$ for $\rho_{\rm lobe}/\rho_0=0.05$, and
$V_{\rm jet} \sim 0.3$ for $\rho_{\rm lobe}/\rho_0=0.1$.
We find these peaks almost correspond to the sound speeds in the plasma sheet,
$c_s = \sqrt{\gamma T_{\rm ps} \rho_{\rm ps}/\rho_0}$,
which are $c_s/V_{A,{\rm lobe}}=0.091$, $0.204$, and $0.289$ for 
$\rho_{\rm lobe}/\rho_0=0.01$, $0.05$, and $0.1$, respectively.
The streaming tearing mode can be amplified in the subsonic jet flow,
while it can be suppressed in the supersonic jet flow.

The reason why the growth rate of the streaming tearing instability 
is enhanced in the subsonic flow may be interpreted from the decrease 
in gas pressure around the X-type point, and the resulting emission of
a fast expansion wave. The streaming plasma around the X-type point can 
be squeezed by the reconnecting magnetic field lines, 
and as a result, the streaming plasma 
speed can increase. Based on Bernoulli's principle of the dynamic and
gas pressure balance along a streamline with $\rho v^2/2 + p=const.$, 
the gas pressure decreases, and the inflow plasma toward the X-type region
can be enhanced. On the other hand, for the case of a supersonic 
flow, the streaming plasma flowing around the squeezed X-type point 
can be decelerated, and then the gas pressure increases. Therefore,
the inflow of plasma toward the X-type point can be suppressed.

A transition from streaming tearing to streaming sausage
modes was observed around $V_{\rm jet}=0.25$, $0.55$, and 
$0.75$ for $\rho_{\rm lobe}/\rho_0=0.01$, $0.05$, and $0.1$, respectively.
The streaming sausage and streaming kink modes are classified as 
nonresistive MHD instabilities, and the mechanism of the instability
may be controlled by the balance between the tension force of the 
magnetic field and the centrifugal force of the bulk flow under an 
infinitesimal distorted magnetic field line. 
If the bulk flow speed is fast enough to overcome the tension force of 
the magnetic field line, then we expect the streaming sausage and streaming kink 
instabilities to occur.
Let us assume that the distortion of the magnetic field lines occurs at the
boundary between the lobe and the plasma sheet at 
$y=\alpha \lambda_{\rm jet}$, 
where $\alpha \sim 1$ is a tuning parameter determined by the details of the 
plasma process. Then, the marginal state of
the instability can be expressed by
\begin{equation}
\frac{B_{\rm lobe}^2 \tanh^2(\alpha')}{4 \pi} = 
  \left(\frac{\rho_{\rm ps}}{\cosh^2(\alpha')}+\rho_{\rm lobe}\right)
  \frac{v_{\rm jet}^2}{\cosh^4(\alpha)},
\end{equation}
where $\alpha'=\alpha \lambda_{\rm jet}/\lambda$. In this case, the onset 
velocity of the streaming sausage may be given by
\begin{equation}
  V_{\rm jet} = \cosh^2(\alpha) \tanh(\alpha')/
     \sqrt{1+\frac{1}{\cosh(\alpha')^2}
           \frac{\rho_{\rm ps}}{\rho_{\rm lobe}}}.
\label{eq:vjet}
\end{equation}
For $\alpha=\alpha'=0.96$ with $\lambda_{\rm jet}=\lambda$, we obtain 
$V_{\rm jet}=0.25$, $0.54$, and $0.75$ for $\rho_{\rm lobe}/\rho_0=0.01$, 
$0.05$, and $0.1$, respectively. This simple estimation can model 
all the transition velocities from the streaming tearing to the streaming sausage modes very 
well.

The transition from the streaming sausage mode to the streaming kink mode, 
which can be recognized by the change in symbol from the red-colored 
circles to the green-colored squares in Figure \ref{fig_LobeN}, 
appears around $V_{\rm jet}=0.4$, $0.8$, and 
$1.05$ for $\rho_{\rm lobe}/\rho_0=0.01$, $0.05$, and $0.1$, respectively.
These transition velocities can be modeled by assuming $\alpha =1.1$ in 
Eq. (\ref{eq:vjet}), and again, we obtain $V_{\rm jet}=0.36$, $0.80$, and 
$1.08$ for $\rho_{\rm lobe}/\rho_0=0.01$, $0.05$, and $0.1$, respectively.

In the earth's magnetotail, the lobe plasma density is lower than 
the plasma sheet density, i.e., 
$\rho_{\rm lobe}/\rho_{\rm ps} \sim 0.01-0.05$ \citep[e.g.][]{Ishisaka01}. 
On the other hand, many MHD simulation studies of magnetic reconnection 
assume a relatively higher plasma density exists in the 
lobe to reduce the Alfv\'en velocity and to save on computational CPU time.
A high density plasma can suppress the emission of the streaming kink 
instability, and this may be one reason why no streaming kink instability 
in association with MHD magnetic reconnection simulations has been reported.

\subsection{Effect of bulk flow size, $\lambda_{\rm jet}$}
\begin{figure}
\centering
\includegraphics[width=20pc]{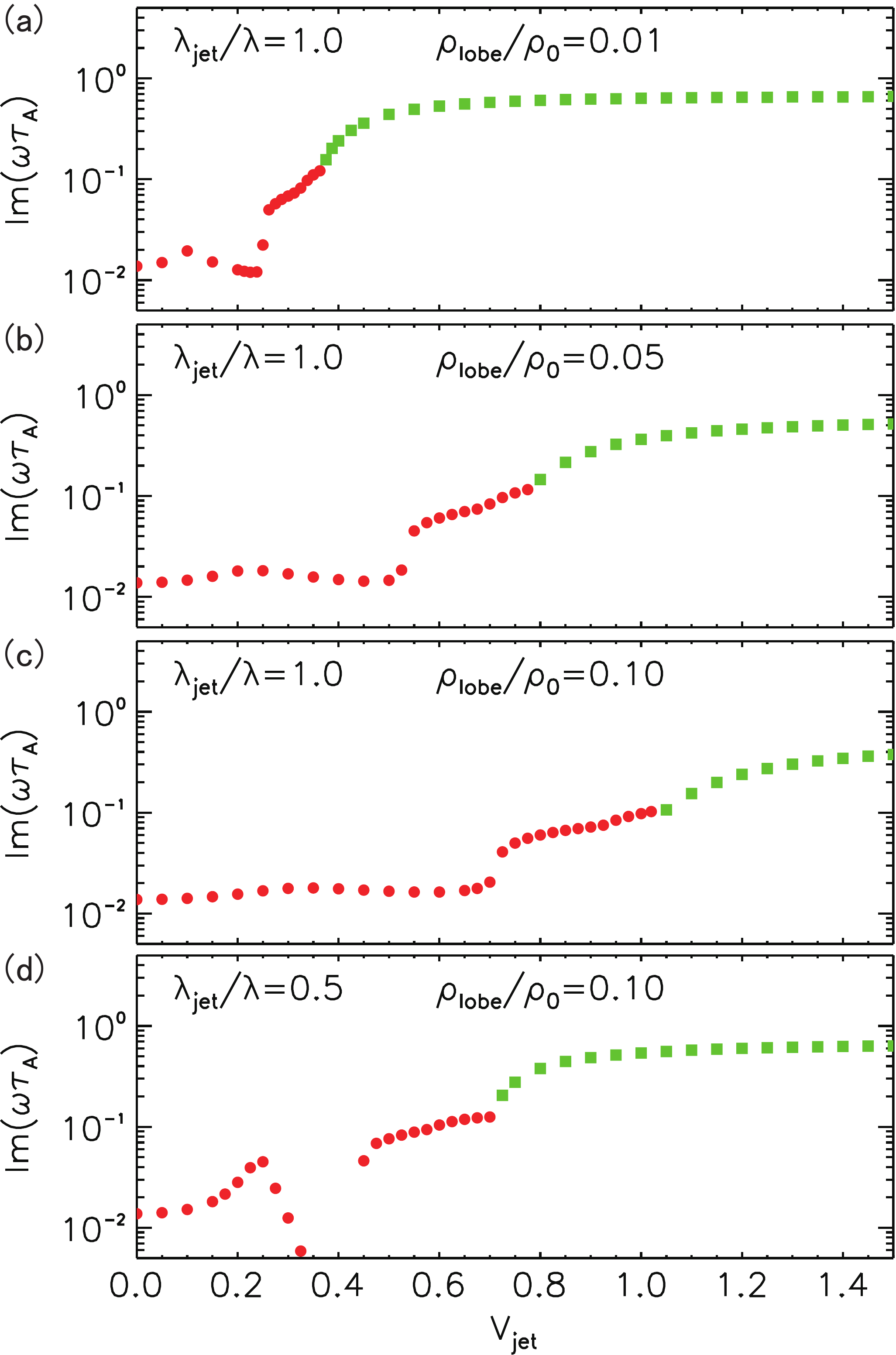}
\caption{Dependence of the lobe plasma density, $\rho_{\rm lobe}/\rho_{\rm ps}$, 
and the bulk flow size, $\lambda_{\rm jet}$.
Maximum growth rates, Im$(\omega \tau_A)$, are shown 
as a function of the bulk flow speed, $V_{\rm jet}$, for the following cases:
(a) $\rho_{\rm lobe}/\rho_0=0.01$; (b) $\rho_{\rm lobe}/\rho_0=0.05$; 
and (c) $\rho_{\rm lobe}/\rho_0=0.1$. (d) The lobe density is the same as 
in case (c), but the thickness of the bulk flow jet is narrower, with 
$\lambda_{\rm jet}/\lambda=0.5$. The red-colored circles and the green-colored squares 
show the symmetric (tearing/sausage) and asymmetric (kink) density 
perturbations, respectively.}
\label{fig_LobeN}
\end{figure}

It is also of interest to study the effect of the jet size 
$\lambda_{\rm jet}$, because the free parameter $\lambda_{\rm jet}$ 
is independent of the equilibrium state.
Figure \ref{fig_LobeN}d shows the result where $\lambda_{\rm jet}=0.5$, 
but keeping $\rho_{\rm lobe}/\rho_0=0.1$, the same as in Figure \ref{fig_LobeN}c. 
By comparing Figures \ref{fig_LobeN}c and \ref{fig_LobeN}d, we find that 
the growth rate of the streaming tearing mode can be enhanced in the 
range $V_{\rm jet} < 0.25$, namely almost in the subsonic flow regime, 
and is reduced in the supersonic flow regime, where the sound speed is 
$c_s/V_{\rm A,lobe}=0.289$. The enhancement and suppression of the
growth rate is much magnified in the case of Figure \ref{fig_LobeN}d.
The reason for this is probably because of the localization of the instability 
inside the high $\beta$ plasma sheet, where the magnetic field is weak. 
That is, for the case of a narrow jet, the instability appears close to the 
neutral sheet with a weak magnetic field, and the effect of the magnetic 
tension/pressure becomes less important compared with the plasma 
dynamic and gas pressures.

The onset velocity of the streaming sausage and streaming kink modes for a narrow jet
is reduced compared with Figure \ref{fig_LobeN}c. This can also be 
interpreted by the weak magnetic field tension force. By using the 
modeling of the onset jet velocities in Eq. (\ref{eq:vjet}) with the 
same values of $\alpha=0.96$ and $\alpha'=0.96(\lambda_{\rm jet}/\lambda)=0.48$,
we obtained $V_{\rm jet}=0.45$, which agrees very well with the onset
of the streaming sausage mode. For the transition from the streaming sausage mode to the
streaming kink mode, by substituting the same value of $\alpha=1.1$ and 
$\alpha'=1.1(\lambda_{\rm jet}/\lambda)=0.55$, we obtained 
$V_{\rm jet}=0.68$. Again, we obtain a good agreement between 
the linear analysis and the theoretical modeling.

The enhancement of the linear growth rate can be understood from 
the localization of the instability in the vicinity of the neutral
sheet because of the narrow jet flow. For a distorted magnetic field line, 
the magnetic tension force becomes weak, while 
the centrifugal force of the bulk flow remains constant, if the bulk
flow speed is constant. Therefore, the linear growth rate can be enhanced 
by reducing the size of the bulk flow. However, since the total free 
energy required to excite the streaming instabilities is reduced, the magnitude of the amplitude 
of the nonlinear saturation of the streaming instabilities may remain small. 
A discussion of the saturation level is beyond the scope of our 
linear analysis.

\subsection{Mode structure under the Hall effect}
So far, we have discussed the behavior of the streaming modes by neglecting the
Hall effect, that is, we have assumed that the ion inertia length is much smaller than
the thickness of the plasma sheet. However, it is known that
the thickness of the plasma sheet is of the order of the ion inertia length 
near the reconnection region. In this subsection, we discuss the effect of the Hall term on
the streaming instabilities.

\begin{figure}
\centering
\includegraphics[width=30pc]{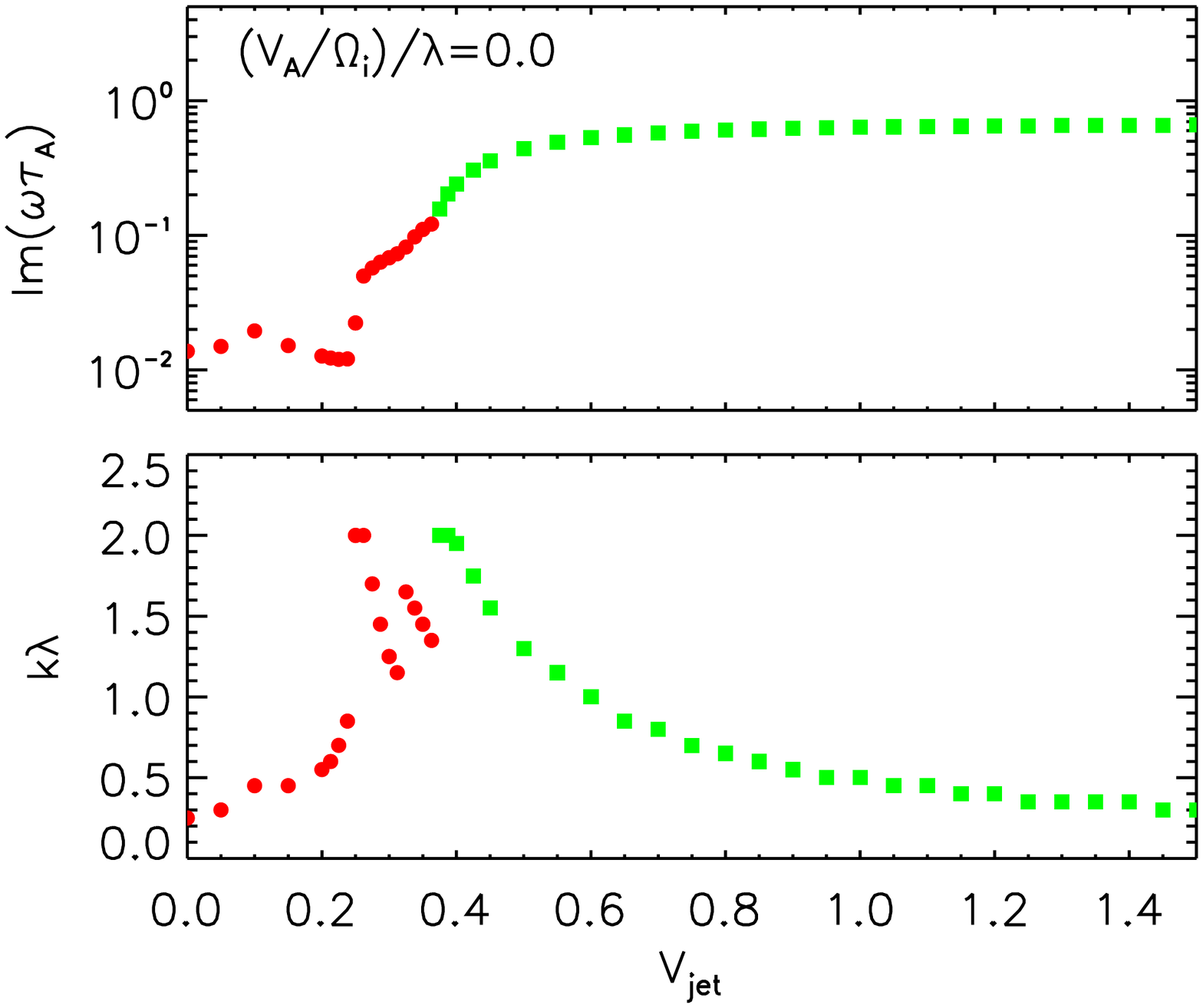}
\caption{Maximum growth rate (top) and the corresponding wave number (bottom) 
as a function of the bulk flow speed, $V_{\rm jet}$, 
for $(V_A/\Omega_i)/\lambda=0$ without a Hall effect. The red-colored circles and
the green-colored squares show the symmetric and asymmetric density perturbations,
respectively.}
\label{fig_Max}
\end{figure}

\begin{figure}
\centering
\includegraphics[width=20pc]{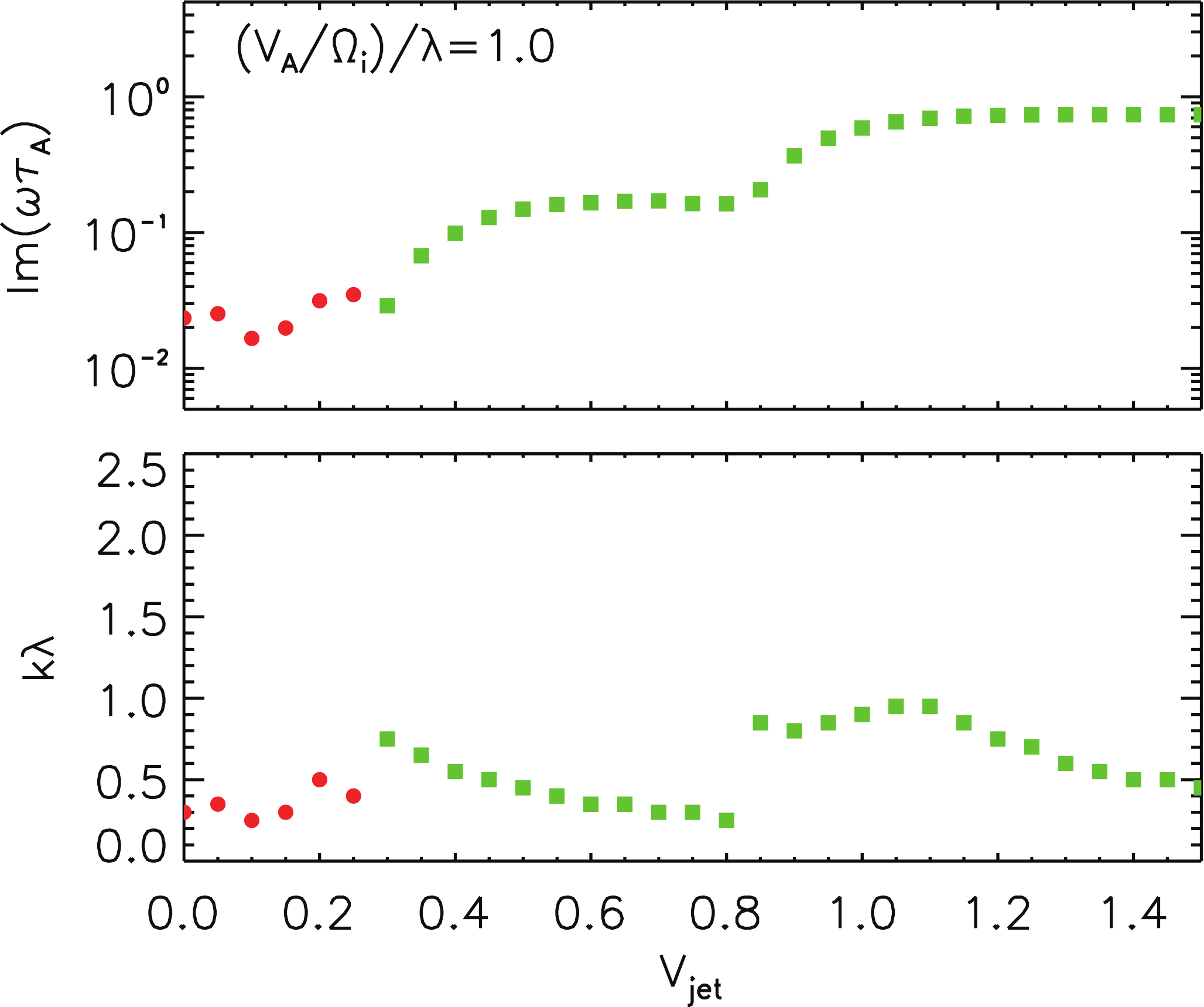}
\caption{Maximum growth rate (top) and its corresponding 
wave number (bottom) are shown as a function of the bulk flow speed, 
under the Hall effect with $(V_A/\Omega_i)/\lambda=1$.  Except for
the Hall parameter, the other parameters are the same as those used 
in Figure \ref{fig_Max}.}
\label{fig_Max_h}
\end{figure}

Figure \ref{fig_Max_h} shows the maximum growth rate and its corresponding 
wave number as a function of the bulk flow speed, $V_{\rm jet}$, 
under the Hall MHD with $(V_A/\Omega_i)/\lambda=1.0$. The other parameters 
are the same as those shown in Figure \ref{fig_Max} without any Hall effect.
By comparing Figures \ref{fig_Max} and \ref{fig_Max_h}, we find that:
(1) the linear growth rate of the Hall tearing mode is larger than that
of the standard MHD tearing mode at $V_{\rm jet}=0$;
(2) regardless of the Hall term, the three different 
streaming tearing, streaming sausage, and streaming kink modes are separated by 
bulk flow speeds of $V_{\rm jet} \sim 0.2$, and $0.3$; and
(3) under a Hall effect, the growth rate does not change much 
but the unstable region is shifted to slightly longer wavelengths.

\begin{figure}
\centering
\includegraphics[width=\textwidth]{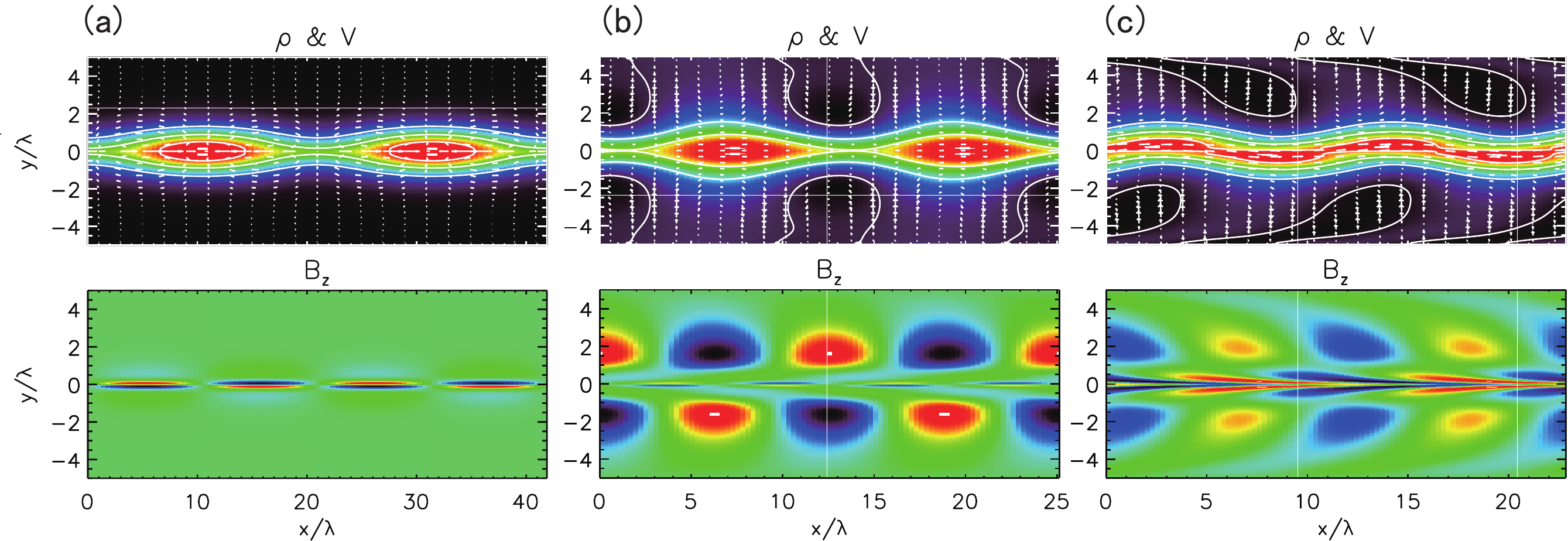}
\caption{Unstable structures for  the plasma density, $\rho$, (top), 
the flow vectors, $\vec{v}$, (white arrows), and the magnetic 
field $B_z$ (bottom): (a) the standard tearing mode without a bulk flow speed $V_{\rm jet}=0$ 
and $k\lambda=0.3$; (b) the streaming sausage mode with $V_{\rm jet}=0.2$ 
and $k\lambda=0.5$; and (c) the streaming kink mode with $V_{\rm jet}=0.4$ 
and $k\lambda=0.55$.}
\label{fig_Egn_h}
\end{figure}

Let us now look at the eigenstructure of the streaming modes under 
the Hall effect shown in Figure \ref{fig_Max_h}. 
Figures \ref{fig_Egn_h}a--c show
the reconstructed structures for $V_{\rm jet}=0$, $0.2$, and $0.4$ 
in the two-dimensional $x$--$y$ space, respectively. 
We have chosen the wave numbers with the maximum growth rate for the fixed
bulk jet speed, namely $k \lambda=0.25$, $0.5$, and $0.5$ for
$V_{\rm jet}=0$, $0.2$, and $0.4$, respectively.
The reconstructed structures were obtained by superposing the first-order 
perturbations onto the zero-order quantities, and we assumed 
the magnitude of the first-order peak density was $30 \%$ of the 
zero-order peak density in the neutral sheet.

In Figure \ref{fig_Egn_h} (a), the quadrupole magnetic field structure 
$B_z$ from the Hall effect can be seen to be localized in the 
vicinity of the neutral sheet. Note that we assumed a magnetic 
Reynolds number of $R_M=10^3$, but a lower value of $R_M$ can produce 
a wider quadrupole magnetic field structure. By increasing the bulk 
flow speed up to (b) $V_{\rm jet}=0.2$ and (c) $V_{\rm jet}=0.4$, 
in addition to the localized $B_z$ component in the vicinity of the 
neutral sheet, large-scale fluctuations outside 
the localized $B_z$ component for both the streaming sausage mode and the 
streaming kink mode were clearly seen. The polarity of $B_z$ is asymmetric for
the streaming sausage mode and symmetric for the streaming kink mode.
In association with the generation of $B_z$, the global plasma flow pattern in the 
entire plasma sheet can be seen, as recognized from the white arrows in the vertical direction.
We believe that the Hall term may have a significant effect on the generation of Alfv\'enic 
fluctuations in the plasma sheet.

\subsection{Guide magnetic field effect}
It is also interesting to study the effect of the guide magnetic field $B_z$ on 
the streaming instabilities, because a global dynamical behavior of magnetic 
reconnection is known to be controlled by the guide magnetic field.  In fact,
the guide magnetic field in the magnetotail, which is the dawn-dusk magnetic field,
is known to exist due to the penetration of the solar wind magnetic field 
\citep[e.g.,][]{Cowley81,Petrukovich11,Rong12}.

\begin{figure}
\centering
\includegraphics[width=25pc]{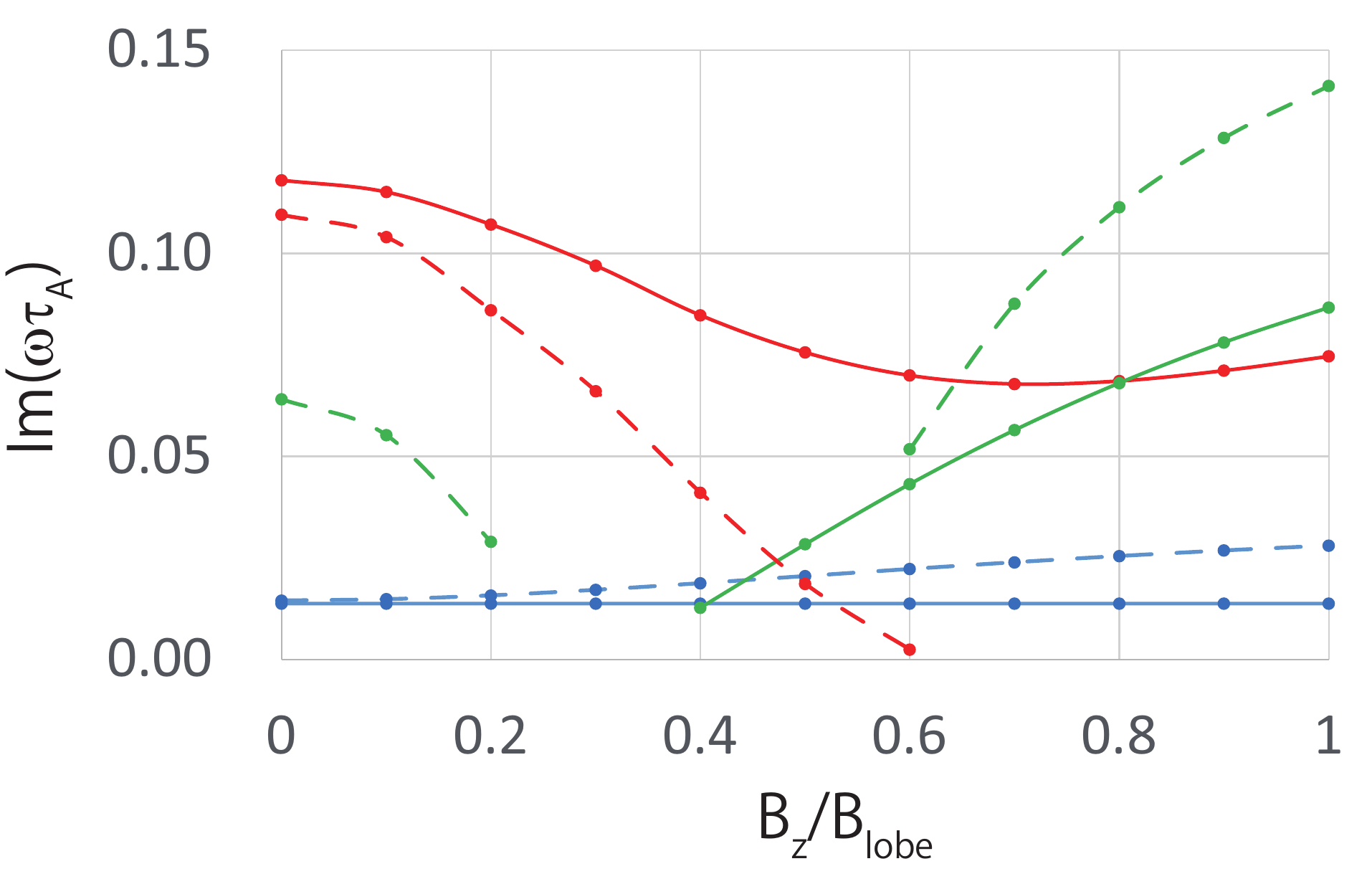}
\caption{Growth rates as a function of the guide magnetic field for
four different unstable modes: 
(a) the standard tearing mode with $(k\lambda, V_{\rm jet})=(0.3,0)$
(blue and solid line), 
(b) the streaming tearing mode with $(k\lambda, V_{\rm jet})=(0.3,0.8)$
(blue and dashed line),
(c) the streaming sausage mode with $(k\lambda, V_{\rm jet})=(1.0,0.8)$ 
(red and solid line) and with $(k\lambda, V_{\rm jet})=(1.5,0.8)$ (red 
and dashed line), and 
(d) the streaming kink mode with $(k\lambda, V_{\rm jet})=(1.0,0.8)$
(green and solid line) and with $(k\lambda, V_{\rm jet})=(1.5,0.8)$
(green and dashed line).}
\label{fig_Bz}
\end{figure}

Shown in Figure \ref{fig_Bz} is the growth rates of the standard tearing mode
and the streaming tearing, sausage and kink modes as a function of 
the initial guide magnetic field $B_z$.  The guide magnetic field is
normalized by the lobe magnetic field $B_{\rm lobe}$, and the growth rate is 
normalized by the Alfv\'en transit time.
The growth rate of the standard tearing mode with $(k \lambda, V_{\rm jet})=(0.3,0)$ 
denoted by the blue-solid line does not depend on the magnitude of the 
guide magnetic field, and that of the streaming tearing mode with 
$(k \lambda, V_{\rm jet})=(0.3,0.8)$ denoted by the blue-dashed line slightly 
increases as increasing the guide magnetic field.
However, the streaming sausage and kink modes with the shorter wavelengths  
show significant change on the guide magnetic field.  The red-solid and red-dashed 
lines show respectively the sausage mode with $(k \lambda, V_{\rm jet})=(1.0,0.8)$ and 
$(k \lambda, V_{\rm jet})=(1.5,0.8)$, while the green-solid and green-dashed
lines are the kink mode with $(k \lambda, V_{\rm jet})=(1.0,0.8)$ and 
$(k \lambda, V_{\rm jet})=(1.5,0.8)$, respectively.  
These sausage and kink modes have a tendency to be 
stabilized for a relatively weak guide magnetic field $B_z/B_{\rm lobe} < 0.5$,
while they are destabilized for a large guide magnetic field $B_z/B_{\rm lobe} > 0.5$.

It should be noted that the inclusion of the guide magnetic field may reduce 
the plasma compressibility, and that the velocity shear instability such as
Kelvin-Helmholtz instability has been shown to be destabilized by a strong
vertical magnetic field perpendicular to the shear plasma flow \citep{Miura82}.
The similarity of the basic behavior between Kelvin-Helmholtz instability 
and the streaming sausage and kink instability has been discussed 
by \citet{Lee88,Biskamp98}, and the incompressible limit due to the 
strong magnetic field may enhance the growth rates of the streaming sausage
and kink instabilities.

\section{Discussion and Conclusions}
We have observed that the streaming MHD instability can be 
activated by not only 
a super-Alf\'enic bulk flow but also a sub-Alf\'enic flow, 
oriented parallel to the antiparallel magnetic field by taking into 
account of the dilute plasma environment in the lobe, 
and found that the growth rates were enhanced more strongly than the 
standard tearing mode without bulk flow. We also studied that the guide 
magnetic field perpendicular to the bulk flow can significantly modify 
the growth rates of the streaming sausage and kink modes.
Furthermore, we also found that Alfv\'enic perturbations of $B_z$
could be generated outside the plasma sheet when the Hall effect
is included. The perturbations may play an important role in the
emission of MHD waves over the plasma sheet. We believe that the 
streaming modes may be a candidate for plasma sheet 
turbulence.  

Since we observed that the bulk flow embedded in the plasma flow
could excite three different streaming modes, depending on the 
bulk flow speed, the next important issue to understand is
the nonlinear evolution and the saturation of the instabilities. 
\citet{Sato82} investigated the 
streaming tearing mode using MHD simulations, and 
suggested that the streaming tearing mode is excited much more 
in the presence of bulk flow. However, 
as \citet{Sato82} assumed that the initial plasma density was uniform 
over the plasma sheet, they only discussed the streaming 
tearing mode with a symmetric density perturbation against 
the neutral sheet. Because a low density plasma in the lobe 
region is required to excite the kink mode, based on our linear 
stability analysis, a streaming kink mode was not demonstrated in 
their simulations. It is important to investigate the nonlinear 
evolution of the streaming kink mode as well. 
Interestingly, some hybrid simulations with a low density plasma in 
the lobe have suggested the presence of a streaming kink-like 
structure in the reconnection jet 
\citep{Lottermoser98,Arzner01,Higashimori12,Liu12}. 
However, since the lobe densities 
in these simulations are still higher than that in the earth's 
magnetotail, we expect that a streaming kink-like structure can be 
excited easily in more realistic situations.

In this paper, we have only discussed the linear stability of the 
streaming MHD instability for an isotropic pressure, i.e., 
$p_\parallel=p_\perp$; however, under a collisionless reconnection, 
it is known that the anisotropy in pressure for 
$p_\parallel > p_{\perp}$ can be generated \citep[e.g.,][]{Hirabayashi13}. 
In this situation, the fire-hose instability may couple with the 
streaming modes \citep{Arzner01}. Under the anisotropic 
plasma situation with  $p_\parallel > p_\perp$,
the magnetic tension force for the distorted magnetic field line
becomes weak; as a result, we may expect that the streaming
instability can be easily excited. However, a study of the
behavior of an anisotropic plasma is beyond the scope of this paper.

We assumed that the physical quantities were uniform in the z direction, i.e.,
$\partial/\partial z=0$, and that the Fourier modes are restricted to be 
propagating parallel to the x direction.  However, it would be interesting to 
understand the oblique propagation as well, because the drifting plasma
population parallel to the electric current may couple with the streaming
plasma flow.  In addition, if there is the density gradient along the plasma 
flow, the interchange instability may be excited in the z direction
\citep[e.g.,][]{Nakamura02,Lapenta11}.  The three dimensional behavior of
the streaming instability would play an important role on the actual plasma sheet.

We have discussed the streaming tearing, streaming sausage, and streaming kink modes as 
possible models of generation of MHD turbulence in the 
earth's magnetotail. However, the streaming instability may be found 
in other applications, one of which is the so-called ``channel flow'' 
formed during the magnetorotational instability (MRI) for 
accretion disks \citep[e.g.,][]{Balbus91}. It is of interest to note that 
the particle-in-cell simulation results on the MRI by \citet{Hoshino13} 
show a streaming kink-type channel flow occurring just before the onset of reconnection. 
The streaming kink mode may also play an important role on the mass and angular 
momentum transport during the formation of an accretion disk.
\begin{acknowledgments}
This work was supported in part by JSPS Grant-in-Aid for Scientific Research 
(KAKENHI) Grant No. 25287151.  The data for this paper are available on request (hoshino@eps.s.u-tokyo.ac.jp).
\end{acknowledgments}
\newpage
\bibliographystyle{agufull08} 

\end{article}
\newpage
\begin{table}
\caption{Plasma parameters used in our linear analysis: 
the normalized lobe plasma density $N_{\rm lobe}=\rho_{\rm lobe}/\rho_0$,
the normalized bulk flow speed by the lobe Alfv\'en speed $V_{\rm jet}$,
the normalized wavelength $k \lambda$, the ratio of the bulk flow jet 
size $\lambda_{\rm jet}$ and the thickness of the plasma sheet $\lambda$, 
the Hall effect on $(V_A/\Omega_i)/\lambda$, and the guide magnetic 
field $B_z/B_{\rm lobe}$.}
\centering
\begin{tabular}{l c c c c c c}
\hline
 Figure  & $N_{\rm lobe}$ & $V_{\rm jet}$ & $k \lambda$ 
         & $\lambda_{\rm jet}/\lambda$ & Hall  & $B_z/B_{\rm lobe}$\\
\hline
  $1a$  & 0.01 & 0.0 & $0 \sim 2$ & 1.0 & 0.0 & 0.0\\
  $1b$  & 0.01 & 0.2 & $0 \sim 2$ & 1.0 & 0.0 & 0.0\\
  $1c$  & 0.01 & 0.4 & $0 \sim 2$ & 1.0 & 0.0 & 0.0\\
  $2a$  & 0.01 & 0.2 & 0.3 & 1.0 & 0.0 & 0.0\\
  $2b$  & 0.01 & 0.4 & 1.5 & 1.0 & 0.0 & 0.0\\
  $3$  & 0.01 & $0 \sim 1.5$ & $0 \sim 3.2$ & 1.0 & 0.0 & 0.0\\
  $4 \rm{(red, solid)}$  & 0.01 & 0.5 & 0.3 & 1.0 & 0.0 & 0.0\\
  $4 \rm{(red, dashed)}$  & 0.01 & 0.5 & 1.0 & 1.0 & 0.0 & 0.0\\
  $4 \rm{(green, solid)}$  & 0.01 & 0.5 & 1.0 & 1.0 & 0.0 & 0.0\\
  $5a$  & 0.01 & $0 \sim 1.5$ & $0 \sim 3.2$ & 1.0 & 0.0 & 0.0\\
  $5b$  & 0.05 & $0 \sim 1.5$ & $0 \sim 3.2$ & 1.0 & 0.0 & 0.0\\
  $5c$  & 0.1 & $0 \sim 1.5$ & $0 \sim 3.2$ & 1.0 & 0.0 & 0.0\\
  $5d$  & 0.1 & $0 \sim 1.5$ & $0 \sim 3.2$ & 0.5 & 0.0 & 0.0\\
  $6$  & 0.01 & $0 \sim 1.5$ & $0 \sim 3.2$ & 1.0 & 1.0 & 0.0\\
  $7a$  & 0.01 & 0.0 & 0.3 & 1.0 & 1.0 & 0.0\\
  $7b$  & 0.01 & 0.2 & 0.5 & 1.0 & 1.0 & 0.0\\
  $7c$  & 0.01 & 0.4 & 0.55 & 1.0 & 1.0 & 0.0\\
  $8a \rm{(blue,solid)}$  & 0.05 & 0.0 & 0.3 & 1.0 & 0.0 & $0 \sim 1$\\
  $8b \rm{(blue,dashed)}$  & 0.05 & 0.8 & 0.3 & 1.0 & 0.0 & $0 \sim 1$\\
  $8c \rm{(red,solid/dashed)}$  & 0.05 & 0.8 & (1.0/1.5) & 1.0 & 0.0 & $0 \sim 1$\\
  $8d \rm{(green,solid/dashed)}$  & 0.05 & 0.8 & (1.0/1.5) & 1.0 & 0.0 & $0 \sim 1$\\
\hline
\end{tabular}
\end{table}
\newpage
\end{document}